\documentclass{aa}
\usepackage{graphics,epsfig,lscape,times}

\def\today{\ifcase\month\or
  January\or February\or March\or April\or May\or June\or
  July\or August\or September\or October\or November\or
  December\fi
  \space\number\day, \number\year}
{\catcode`\@=11
\gdef\SchlangeUnter#1#2{\lower2pt\vbox{\baselineskip 0pt\lineskip0pt
   \ialign{$\m@th#1\hfil##\hfil$\crcr#2\crcr\sim\crcr}}}}

\begin{document}
\def\gray{$\gamma$-ray\ }
\def\grays{$\gamma$-rays\ }
\def\phibar{$\bar\varphi$\ }
\def\deg   {$^o$\ }
\def\intfluxunits{\  cm$^{-2}$s$^{-1}$\ }
  \def\cms{cm$^{-2}$ s$^{-1}$}
  \def \ti{$^{44}$Ti }
  \def \sc{$^{44}$Sc }
  \def \ca{$^{44}$Ca }
  \def \k{$^{40}$K }
  \def \msol {M$_{\odot}$\ }
  \def \about{$\sim$ }
  \def \sig{$\sigma$ }
  \def \c {$^{12}$C }
  \def \o {$^{16}$O }
  \def \ni{$^{56}$Ni }
  \def \fe{$^{56}$Fe }
  \def \co{$^{56}$Co }
  \def \na{$^{22}$Na }
  \def \gam{$\gamma$ }
  \def \cms{cm$^{-2}$ s$^{-1}$ }
  \def \lb{\hfil\break}

\title{XMM-Newton observations of the supernova remnant \\
RX J0852.0-4622 / GRO J0852-4642}

\author{A. F. Iyudin, B. Aschenbach, W. Becker, K. Dennerl, F. Haberl}
 \offprints{A. F. Iyudin\\ (aiyudin@srd.sinp.msu.ru)}

\institute{Max--Planck--Institut f\"ur extraterrestrische Physik, 
           85741 Garching, Germany 
          }
\date{Received ...;   Accepted ... }

\abstract{
RX J0852.0-4622 is a supernova remnant discovered in the {\it{ROSAT}} all-sky survey. Spatially coincident 
1.157 MeV $\gamma$-ray line emission was detected with the {\it{COMPTEL}} instrument on-board of the 
CGRO. The analysis combining the X-ray and $\gamma$-ray data suggests that  RX J0852.0-4622 is a close-by and young 
supernova remnant. Follow-up observations with {\it{ASCA}} show that the two brightest sections of the limb have 
non-thermal spectra, which make an independent assessment of the age and distance using the Sedov equations 
for the evolution of the remnant almost impossible. We have observed three rim sections of RX J0852.0-4622 with 
{\it{XMM-Newton}} and confirm the power law type spectra measured with {\it{ASCA}}. We also confirm 
the presence of an emission line like feature at 4.45 $\pm$ 0.05 keV, which we suggest to be emission from  Ti and Sc 
excited by atom/ion or ion/ion high velocity collisions. The high velocity is in agreement with the 
width of the 1.157 MeV $\gamma$-ray line. The X-ray line flux expected from such an interaction is consistent with 
the 1.157 MeV $\gamma$-ray line flux measured
by {\it{COMPTEL}}. This consistency of the X-ray line flux and the $\gamma$-ray line flux
lends further support to the existence and amounts of Ti in RX J0852.0-4622 claimed by Iyudin et al. (1998) and to the suggestion
that RX J0852.0-4622 is young and nearby (Aschenbach et al. 1999).
Iyudin et al. (1998) quote a very large broadening of the 1.157 MeV $\gamma$-ray line which would indicate a large
velocity of the emitting matter of about 15.000 km/s. Such high ejecta velocity for Ti is found only in explosion models of
sub-Chandrasekhar type Ia supernovae (Woosley \& Weaver 1994, Livne \& Arnett 1995).
  In this case no compact remnant is expected. The obvious questions remaining are what the nature and the origin of the
central compact source CXOU J085201.4-461753 are and why the absorption column density apparently associated with RX J0852.0-4622 is much
greater than the typical column for the Vela SNR.
}
\titlerunning{{\it{XMM-Newton}} observations of RX J0852.0-4622}
\authorrunning{Iyudin et al.}
\maketitle
%

%
\section {Introduction}
The {\it{ROSAT}} all-sky survey has revealed a number of previously unknown 
supernova remnants (SNR) among which is  \object{RX J0852.0-4622}, sometimes 
referred to as \object{G266.2-1.2}, which is located at the 
south-eastern corner of the Vela SNR (Aschenbach 1998). 
The all-sky map of $\sim$6 years of {\it{COMPTEL}} data shows 
a few localized 1.157 MeV  line emission features. 
This $\gamma$-ray line originates from the 
decay of 
radioactive $^{44}$Ti (Iyudin et al. 1998, Iyudin 1999). 
 The most significant
excess is attributed to the \object{Cas A} SNR whereas the second brightest structure called 
\object{GRO J0852-4642} coincides with 
RX J0852.0-4622 (Iyudin 1999, Iyudin et al. 1998). Since $^{44}$Ti is exclusively produced in supernovae it is very likely 
that RX J0852.0-4622 and GRO J0852-4642 are the same object which was created in one supernova explosion. 
The combined analysis of the X-ray data and the $\gamma$-ray data led to the suggestion that RX J0852.0-4622 could be 
the remnant of the nearest supernova in recent history  (Aschenbach et al. 1999) with a best estimate for the distance of 
200 pc and an age of 680 years. The detection of the 1.157 MeV $\gamma$-ray line from radioactive
$^{44}$Ti in Cas A (Iyudin et al. 1994, 1997) has been supported by the {\it{Beppo-SAX}} measurements of the $\sim$68
and $\sim$78 keV X-ray lines (Vink et al. 2001) which are produced in the first decay of $^{44}$Ti in the
decay chain $^{44}$Ti$\rightarrow$$^{44}$Sc$\rightarrow$$^{44}$Ca. The detection of the 1.157
MeV $\gamma$-ray line from Cas A was the first discovery of $^{44}$Ti in a
young galactic SNR, and as such it provides an
essential calibration of nucleosynthesis model calculations. 
With the discovery of RX J0852.0-4622/GRO J0852-4642 we may have a second example. 

\par
Follow-up observations in the radio band confirmed the shell-like structure and supported the identification 
of RX J0852.0-4622 as a SNR (Combi et al. 1999, Duncan and Green 2000, Filipovic 2001). Furthermore a good correlation 
between X-ray and radio brightness was found (c.f. Fig. 1, Filipovic 2001). 
\begin{figure}[hbt]
  \centering
\includegraphics[bb=70 370 532 745,width=8.5cm,clip]{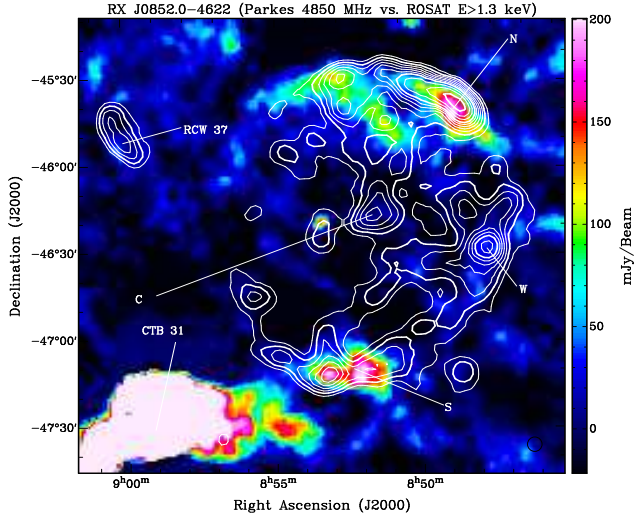}
  \caption{The Parkes image (colour) of RX J0852.0-4622 overlaid with ROSAT PSPC contours (E$>$1.3 keV). 
The synthesized beam of the Parkes observations is 5$'$ (lower left corner) with an r.m.s noise (1 $\sigma$)
of $~$10 mJy. X-ray contours (white) are 0.8, 1, 1.2, 1.4, 1.6, 1.8, 2, 2.3, 2.6 and 3 in units of 10$^{-4}$
ROSAT PSPC counts s$^{-1}$ arcmin$^{-2}$ (Filipovic 2001).}
\end{figure}
\par
The spectra taken with {\it{ROSAT}} are compatible with either emission from a high temperature plasma or a non-thermal source 
creating a power law 
(Aschenbach et al. 1999). The X-ray surface brightness of RX J0852.0-4622 is generally low and implies a rather low
matter density for any shock wave heated plasma.
Follow up measurements with {\it{ASCA}} in December 1998 generally confirmed the results of {\it{ROSAT}} but demonstrated that
the spectrum is definitely non-thermal with no obvious emission lines above 1 keV (Tsunemi et al. 2000, Slane et al. 2001).
The best-fit with a single power law results in an interstellar absorption column density which is at least a factor of three higher 
than the highest value observed elsewhere in the \object{Vela SNR} (Lu \& Aschenbach 2000). 
\par
Based on the {\it{ROSAT}} data
Aschenbach (1998) suggested the presence of a central point source, which later was confirmed
by {\it{Chandra}} measurements (Pavlov et al.
2001). The spectrum of the source  \object{CXOU J085201.4-461753} suggests a neutron star as the emitter (Kargaltsev et al. 2002), which is supported 
by the absence of any optical counterpart brighter than B$\sim$23 (Mereghetti et al. 2002).
If this object is the compact remnant of the supernova which created RX J0852.0-4622 the supernova was of the 
core-collapse type. Like for the north-western rim the absorbing column density inferred from the spectrum of CXOU J085201.4-461753 is significantly higher
by a factor of about six (Kargaltsev et al. 2002) than the column densities measured for the Vela SNR (Lu \& Aschenbach 2000),
which suggests that at least CXOU J085201.4-461753 and possibly RX J0852.0-4622 are at a greater distance than
the Vela SNR, which is located at a distance of about 250 pc (Cha et al., 1999).
On the other hand, Redman et al. (2002) suggested, as one possibility, that the optical nebula 
\object{RCW 37} was generated
by the blast wave of RX J0852.0-4622 impacting the shell of the Vela SNR, i.e. RX J0852.0-4622 is embedded in the Vela SNR.
From observations of 60 O stars  Cha \& Sembach (2000) have recommended a distance of 390$\pm$100 pc to the Vela SNR.
Along the line of sight towards the O star \object{HD 75821} (BO IIIFs at l, b = 266$\fdg$.25 / -1$\fdg$.54) at a distance of $\sim$830$\pm$360 pc absorption
lines of Ca II at 3933 \AA~ and of Na I at 5889 \AA~ were found. This line of sight is passing
 through  the RX J0852.0-4622 shell near its center.
From these observations a distance limit to RX J0852.0-4622 of $\leq$830$\pm$360 pc can be derived.
Pozzo et al. (2000) conclude that the distance to RX J0852.0-4622 is $\approx$430$\pm$60 pc.
From radio observations of the CS J = 2 - 1 transitions towards the
source \object{G 267.94-1.06}, which is  at a distance of 700 pc,  a cloud was found which has a core mass of 1500 $M_{\odot}$ and an extent of
 $\sim$1.4 pc along the line of sight, or a mean density of 10$^4$ cm$^{-3}$ of CS (Zinchenko, Mattila and Toriseva, 1995).
Deep absorption features in the CO and CS profiles were observed in several sources clustering
mainly around $l$$\approx$270$\degr$. The authors argue that this can be attributed to extended low-excitation
foreground clouds. It means that the  distance to the south-eastern rim of RX J0852.0-4622 can be $<$700 pc.
From the observations of the \object{Gum Nebula} in the OH line at 1667 MHz
the distance to the \object{Vela OB2 association} was derived to 420$\pm$30 pc (Woermann et al. 2001), which would also be the distance of
RX J0852.0-4622 if it would belong to Vela OB2. 
\par
Concerning the age, there has been the suggestion, that the supernova which led to RX J0852.0-4622 was responsible 
for a previously unidentified spike in nitrate concentration measured in an Antarctic ice core. The precipitation 
occurred around the year 1320. Other nitrate spikes could be associated with historical 
supernovae (Burgess \& Zuber 2000).  
One of the issues in this context is of course the detection of the 1.157 MeV line, because together with the
$^{44}$Ti yield it dominates the estimate of the age.
There have been numerous discussions about the significance of the 1.157 MeV line (see e.g.
 Aschenbach et al. 1999, Iyudin 2000,
Sch\"onfelder et al. 2000, Iyudin and Aschenbach 2001) and it is obvious that an independent confirmation is needed.
First attempts made with {\it{Integral}} have so far produced an upper limit, which is about three times as high as the
flux measured with {\it{COMPTEL}} (von Kienlin et al. 2004).

\par
A totally different explanation for RXJ0852.0-4622 has been given by Wang \& Chevalier (2002), who suggest that the ring-like 
radio and X-ray structure could have been produced by clumps generated in and by the Vela SNR. 
To shed further light on the questions including distance, age and progenitor we carried out {\it{XMM-Newton}} observations 
of three circular sections distributed along the rim of RX J0852.0-4622 (Fig. 2). The results obtained from the measurements of the 
central region including the compact source CXOU J085201.4-461753 will be reported elsewhere. 
The earlier {\it{ASCA}} observations were dedicated to study the enhanced X-ray emission regions along the northern and
western limbs, the results of which we compare with our findings. We have added an observation of the   
south-eastern X-ray bright shell section which was not 
observed by {\it{ASCA}} (Tsunemi et al. 2000, Slane et al. 2001). Generally    
the {\it{ASCA}} observations showed that the spectra follow a power law with a photon spectral index of $\sim$2.6 and 
no emission lines have been detected in the GIS spectra (Tsunemi et al. 2000, Slane et al. 2001). 
But the analysis of the spectra taken with the SIS 
reveals a line feature at an energy
of $\sim$4.1 keV in the spectrum of the north-western (NW) rim (Tsunemi et al. 2000).
The best-fit model actually consists 
of two thermal components and an emission line 
at 4.1$\pm$0.2 keV. This model implies a large overabundance of Ca based on the line flux at 
 4.1 keV, which was interpreted as the He-like
K emission line of $^{44}$Ca, which is the final product of the decay of $^{44}$Ti (Tsunemi et al. 2000). 
A feature at 4.1 keV in the SIS0 chip of the {\it{ASCA}} detector was also noted by Slane et al. (2001), but the authors 
were not convinced of its presence because of the low significance. However, if interpreted as the $^{44}$Sc-K fluorescence line, the immediate 
decay product of $^{44}$Ti, the upper limit of the Sc flux like the Ca flux is not inconsistent with the flux expected from the $^{44}$Ti 
flux measured by Iyudin et al. (1998).   
If the X-ray lines are present at a flux level proposed by the observations the age of  RXJ0852.0-4622 is not inconsistent 
with $\sim$1000 years (Tsunemi et al. 2000). 
Previously  we have suggested (see Iyudin \& Aschenbach, 2001) that a joint analysis of the $\gamma$-ray
line emission data from radioactive elements in combination
with the spatially resolved X-ray spectroscopic data 
can provide stringent constraints on the abundance of some elements and on the type of the supernova.
Here we compare results of the
$\gamma$-ray data analysis of the
SNR RXJ0852.0-4622/GROJ0852-4642 obtained with {\it{COMPTEL}} on-board of the Compton Gamma-Ray Observatory (CGRO)
with the spatially resolved X-ray spectra obtained with {\it{XMM-Newton}}.
Specifically, we discuss the X-ray continuum shape at $E_X$$\geq$1.0 keV along the X-ray bright rim of the
SNR, and the line feature at $\sim$4.2 keV. We believe that the X-ray line at $\sim$4.2 keV is a direct
consequence of the $^{44}$Ti decay in SNR shell, which would imply the presence of high-velocity $^{44}$Ti in
RX J0852.0-4622, and which is suggested by the width of the 1.157 MeV $\gamma$-ray line first measured with {\it{COMPTEL}} (Iyudin et al. 1998).

\section{Observations}  

The diameter of RX J0852.0-4622 is $\sim$2\deg which is significantly larger than the $\sim$$30'$ field of view of the EPIC instruments 
on board of {\it{XMM-Newton}}. Four different pointings on the brightest sections of the remnant were carried out in the GTO program, three 
of which were directed to the rim, i.e. the northwest (NW), the west (W) and the south (S) and the fourth pointing was towards the 
center (C) (Fig. 2). The observations were carried out between April 24 and April 27, 2001. The EPIC-PN camera (Str\"uder et al. 2001) was operated in 
extended full frame mode and the medium filter was in place. The EPIC-MOS1 and -MOS2 cameras (Turner et al. 2001) were used in full frame mode with the medium 
filter as well. Further details of the observations are given in Table 1. 

\begin{figure}[hbt]
  \centering
\includegraphics[bb=75 315 530 770,width=8.5cm,clip]{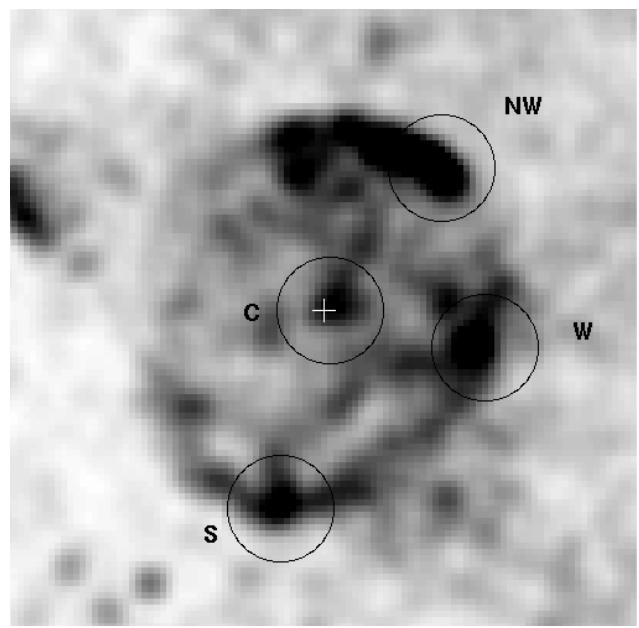}
  \caption{ROSAT grey scale image of RX J0852.0-4622 for E$_x$$\geq$1.3 keV;
black circles show the {\it{XMM-Newton}} GTO fields in the northwest (NW), the west (W), 
the south (S) and the center (C) of the remnant. North is up and east is to the left. The white cross marks the position 
of the point source CXOU J085201.4-461753.}
\end{figure}

\begin{table*}[hbt]
\caption{{\it{XMM-Newton}} GTO observations of RX J0852.0-4622}

\smallskip
\begin{center}
\begin{tabular*}{78mm}{||c|cc|c|c||@{\extracolsep\fill}} \hline
  &\multicolumn{2}{|c|} {Pointing (J2000)}  & Expos.& XMM \\
\cline{2-5}
  \raisebox{1.5ex}[0pt]{Rim} & RA & DEC & ksec & revol.  \\
\hline
\hline
NW & 08h48m58s  & -45d39m03s  & 31.76 &  0252  \\
\hline
West& 08h47m45s  & -46d28m51s  & 35.96 &  0252  \\
\hline
South & 08h53m14s  & -47d13m53s  & 47.13 &  0253  \\
\hline
Center& 08h51m50s  & -46d18m45s  & 24.11 &  0252  \\
\hline
\hline
\end{tabular*}
\end{center}

\par
\end{table*}

\section{Spectral fits}

RX J0852.0-4622 is located in the south-eastern corner of the Vela SNR, which actually completely covers RX J0852.0-4622. At low energies 
the X-ray surface brightness of the Vela SNR is much higher than that of RX J0852.0-4622, so that RX J0852.0-4622 becomes visible only 
above $\sim$1 keV in the {\it{ROSAT}} images. This is a major complication for any spectral fits to RX J0852.0-4622. We have followed three options. 
In a first approach we chose fields for the background to be subtracted  which are definitely outside the area covered by RX J0852.0-4622 but inside 
the Vela SNR. The corresponding fits did not converge to a unique solution, when different fields for the background were chosen. This is likely 
to be due to the Vela spectra changing on a scale of a few arcminutes (Lu \& Aschenbach 2000). In a second approach we fit the spectra of RX J0852.0-4622 with a three component model. This model consists of one thermal component with an associated 
interstellar absorption column density which together represent the low energy Vela SNR emission; the second component is again thermal emission with an added 
power law (third component) to represent the higher energy emission. The second and third components have the same absorption column density. This model was applied 
to the NW rim and the S rim, and the results are shown in Table 2. 

\begin{center}
\begin{table*}[bht]
\caption{Spectral fit results with a {\it wabs(vmekal)+wabs(vmekal+power law)} model for the range 0.2$\leq$E$_x$$\leq$10 keV.}

\begin{tabular*}{154mm}{||l||c|c|c|c|c|c|c||@{\extracolsep\fill}} \hline
   & low kT$_1$ & high kT$_2$ & $N_H$(kT$_1$) &$N_H$(kT$_2$) & power law &  $\chi$$^2$ (dof) & ${\chi}^2_{\nu}$  \\

 \raisebox{1.5ex}[0pt]{Region} & (eV) & (keV) &(10$^{22}$ cm$^{-2}$) &(10$^{22}$ cm$^{-2}$) & photon index&  & \\
\hline
\hline
 NW rim  & 37.5$^{+4.0}_{-2.0}$ & 3.84$^{+1.65}_{-1.30}$  & 0.138$^{+0.022}_{-0.026}$ & 0.461$^{+0.023}_{-0.019}$ & 2.59$^{+0.16}_{-0.07}$ & 796.48 (777) & 1.025 \\
\hline
 Southern rim 1  & 43.5$\pm$0.8 & 2.92$\pm$2.50  & 0.123$\pm$0.016 & 0.471$\pm$0.035 & 2.55$\pm$0.25 & 349.01 (379) & 0.92 \\
\hline
 \hline
\end{tabular*}
\end{table*}
\end{center}       

As expected from the work of Lu \& Aschenbach (2000) there is a low temperature component 
({\it{vmekal1}}) with a temperature between 37 and 44 eV associated 
with a column density around 1.3$\times$10$\sp{21}$ cm$\sp{-2}$, which is about a factor of two to three higher than found by Lu \& Aschenbach 
(2000). But the column density for the high temperature ({\it{vmekal2}})/ power law components is higher by another factor of three to four, which would indicate 
a larger distance if the spectral model is correct. Formally, i.e. based on ${\chi}^2_{\nu}$  (c.f. Table 2), the fits are acceptable for both 
the NW and S region, with no significant difference of the best fit parameters for the two regions. 
The power law slopes of 2.59 and 2.55, respectively, and the absorption column 
densities agree remarkably well with the {\it{ASCA}} measurements (Tsunemi et al. 2000, Slane et al. 2001) 
despite the significantly higher photon statistics and higher energies covered by {\it{XMM-Newton}}. 

Because of the ambiguity concerning the contribution of the Vela SNR we restricted in what follows the energy range for 
the fits to E$>$0.8 keV, which is our third approach to model the spectra.

\subsection{The north-western rim}

Figure 3 shows the {\it{XMM-Newton}} EPIC-PN image of the north-western rim. A bright filament like structure defines the outer edge of the 
remnant and a second, non-aligned and significantly fainter filament like structure is seen further inside. The two structures seem to join each other 
at the north-eastern tip. This is actually the first X-ray image which resolves the remnant's outer boundary. 

\begin{figure}[hbt]
  \centering
\includegraphics[bb=68 205 510 601,width=8.5cm,clip]{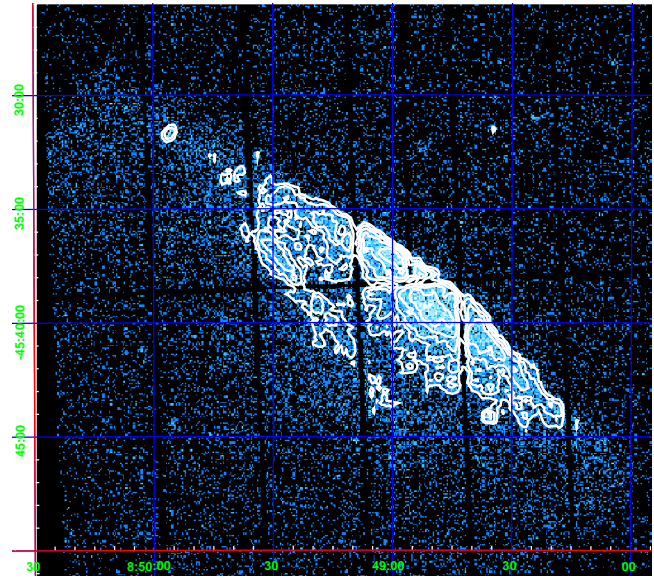}
  \caption{The {\it{XMM-Newton}} EPIC-PN image of the bright NW part of the RX J0852.0-4622 shell, which is spatially resolved 
            for the first time.}
\end{figure}

In an attempt to improve upon the non-thermal component modelling of the spectrum with  a straight power law model, we have 
tried in addition the
synchrotron model ({\it sresc}) of XSPEC, developed by Reynolds (1996, 1998). This model represents synchrotron
emission from the shock wave accelerated high-energy electrons and which takes care of  
electron escape or synchrotron losses 
by a steepening of the spectrum towards higher 
energies. 
As can be seen from Table 3 the reduced ${\chi}^2_{\nu}$ value is slightly lower than for a 
straight power law. As Figure 4 shows two local residuals around $\sim$4.4 keV and $\sim$6.5 keV are left from the fit.  
By adding one or two Gaussian  shaped line(s) {\it{gauss}} at these energies to the {\it sresc} continuum
model it is possible to further lower ${\chi}^2_{\nu}$ (Table 3). We note that a simple power law model is quite
acceptable judging on the total ${\chi}^2$ value. We also note that we still need a thermal component to cover the low energies 
which is represented by a {\it{vmekal}} model

\begin{figure}[hbt]
  \centering 
\includegraphics[bb=70 460 532 770,width=8.5cm,clip]{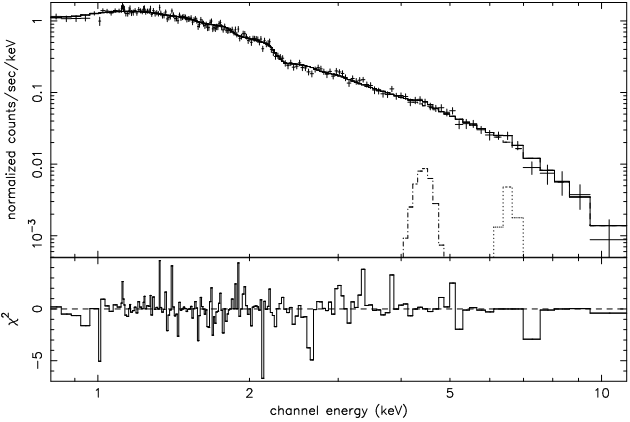}
  \caption{PN spectrum of the brightest region of the north-western rim of RX J0852.0-4622 for E$>$ 0.8 keV in comparison with a {\it sresc+2 gauss} model fit.}
\end{figure}

\begin{center}
\begin{table*}[bht]
\caption{Fit results for E$_x$$\geq$0.8 keV with various spectral models for the continuum including a power law, the {\it sresc} model with and without 
one or two Gaussian lines.}

\begin{tabular*}{142mm}{||l||l||c|c|c|c|c||@{\extracolsep\fill}} \hline
 &  & ph. index, or &$N_H$ & radio index &  $\chi$$^2$ (dof) & ${{\chi}^2}_{\nu}$  \\

 \raisebox{1.5ex}[0pt]{Region} &  \raisebox{1.5ex}[0pt]{Model}& ${\nu}_{rolloff}$, Hz &(10$^{22}$ cm$^{-2}$) & $\alpha$ at 1 GHz&  & \\
\hline
\hline
 NW rim  & {\it powerlaw} & 2.60  & 0.496 & --- & 166.64 (173) & 0.963 \\
\hline
 NW rim  & {\it sresc} & (2.2$^{+0.4}_{-0.2}$)$\times$10$^{17}$  & 0.389 & 0.24 & 157.76 (172) &  0.917 \\
\hline
 NW rim  & {\it sresc+gauss} & 2.20$\times$10$^{17}$  & 0.389 & 0.24 & 147.06 (170) & 0.865 \\
\hline
 NW rim  & {\it sresc+2gauss} & 2.20$\times$10$^{17}$  & 0.389 & 0.24 & 145.04 (170) & 0.853 \\
\hline
 Southern rim  & {\it sresc} & (2.6$^{+0.6}_{-0.4}$)$\times$10$^{17}$  & 0.414 & 0.31 & 230.707 (225) & 1.025 \\
\hline
 \hline
\end{tabular*}
\end{table*}
\end{center}

\begin{figure*}[hbt]
  \centering
\includegraphics[bb=20 495 575 682,width=17.5cm,clip]{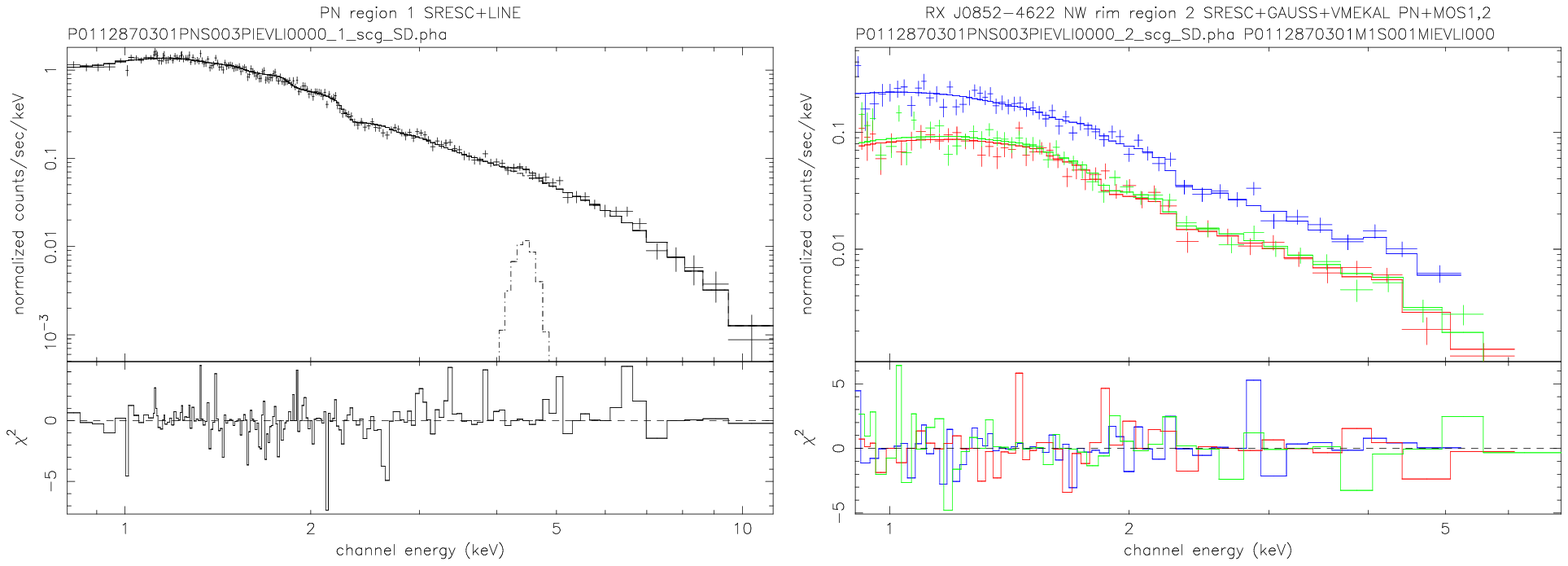}
  \caption{Left: PN spectrum of the brightest region of the RX J0852.0-4622 NW rim (region 1). Right: EPIC-PN and EPIC-MOS1,2
spectra of fainter filament of the NW rim (region 2). Spectra were fitted for E$>$ 0.8 keV with the {\it vmekal+sresc+gauss} model.}
\end{figure*}
We have also analysed the X-ray spectrum of each of the two filaments (c.f. Fig. 3) separately. 
Both filaments have a power law continuum spectrum
for E$_x$$\geq$0.8 keV, which can be also well fitted by the {\it{sresc}} model of XSPEC 
(c.f. Fig. 5). The line feature at $\sim$4.2 keV
detected by {\it{ASCA}} (Tsunemi et al. 2000), is also detected by {\it{XMM-Newton}} (Iyudin et al. 2004), and is present in both filaments.
The line flux is larger in the brighter (outer) filament. Unfortunately, low statistics preclude a reasonable
analysis of the line flux radial dependence. We believe that the X-ray line at $\sim$4.2 keV is a direct
consequence of the $^{44}$Ti decay in the SNR shell. Note the sharp outer boundary of the remnant and the
clear presence of two arc-like features. The fainter one might correspond to the reverse shock, that only starts to develop.
We have detected two line-like excesses, one of which is also visible in the spectrum of
the southern rim (see below).
The strongest line is at E$_{line}$=4.24$^{+0.18}_{-0.14}$ keV for the NW reg. 1 for a power law continuum model, and
E$_{line}$=4.44$\pm$0.11 keV for the same region using the {\it sresc} model. Our values of the line energy are slightly higher
than the value given by Tsunemi et al. (2000), but are consistent with the {\it{ASCA}} SIS spectrum of the NW rim (Fig. 4 in
the paper of Tsunemi et al. 2000).
\par
Assuming that the line feature is a blend of the hydrogen- and He-like lines of Ca, and that all the Ca consists
of $^{44}$Ca, we expected to detect a flux of $\sim$5.6$\times$10$^{-6}$ ph cm$^{-2}$ s$^{-1}$ from the whole SNR RX J0852.0-4622,
based on the flux of (3.8$\pm$0.7)$\times$10$^{-5}$ ph cm$^{-2}$ s$^{-1}$ of the line at 1.157 MeV from
the $^{44}$Ti decay (Iyudin et al. 1998).
The significance of the line detection taking exposure and background into account corresponds to $\sim$4 $\sigma$ for
the whole NW rim. A higher significance is actually not expected given the 1.157 MeV flux quoted above. The detection
of a line with {\it{ASCA}} at $\sim$4.1 keV is therefore confirmed by the {\it{XMM-Newton}} GTO measurements, but the significance is
still inconclusive to claim the existence of the line beyond any doubt, although the near coincidence of the line
energies is rather compelling.
   
\subsection{The southern rim}

The EPIC-MOS1 image of the southern rim of RX J0852.0-4622 is shown in Figure 6. 

\begin{figure}[hbt]
  \centering
\includegraphics[bb=215 575 383 708,width=8.0cm,clip]{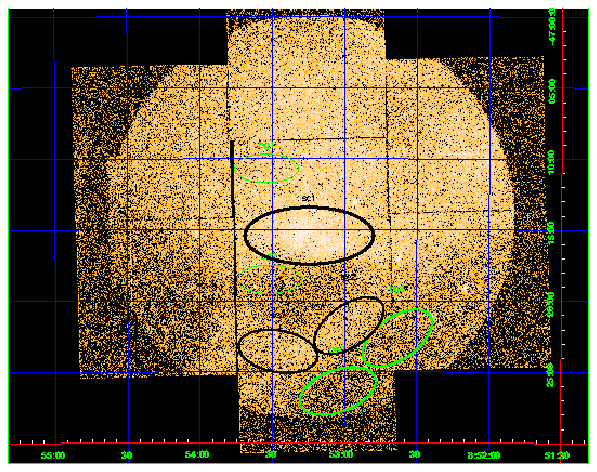}
  \caption{The {\it{XMM-Newton}} EPIC-MOS image of the bright southern part of the RX J0852.0-4622 shell. The black 
ellipse shows the extraction region for the  source spectrum. Green ellipses
show the regions from where the background spectra were taken. }
\end{figure}

PN and MOS1,2 spectra have been created for the counts of the region 
enclosed by the large black ellipse (region 1) and are shown in Figure 7. Again, an emission line like excess 
at $\sim$4.2 keV appears in the spectra. 
\begin{figure*}[hbt]
  \centering
\includegraphics[bb=20 495 575 682,width=17.5cm,clip]{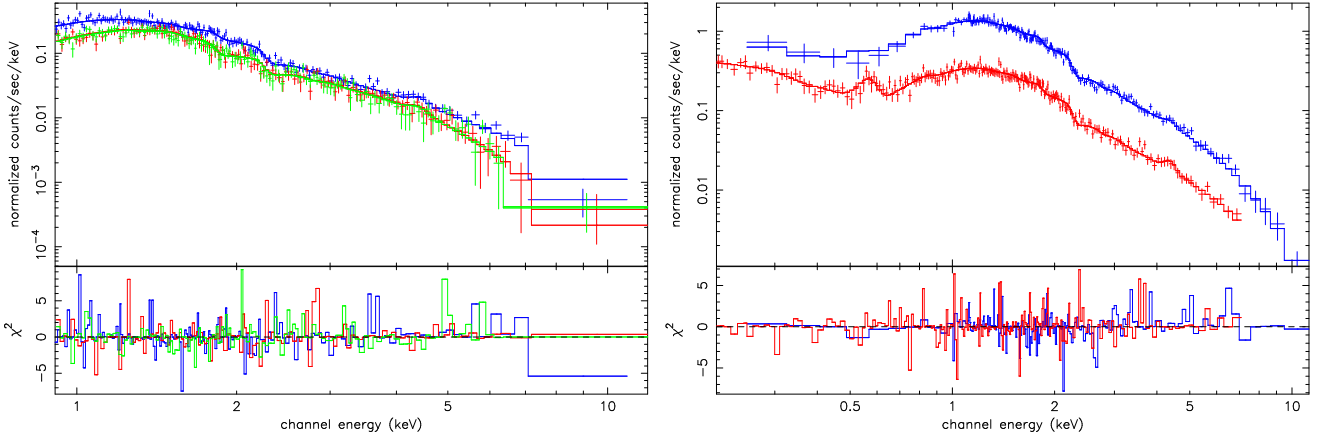}
  \caption{ Left: EPIC-PN and EPIC-MOS1,2 spectra of the RX J0852.0-4622 southern rim brightest region 1 
for E$>$ 0.8 keV. Right: EPIC-PN spectra of the north-western rim (top curve) and of the southern rim 
(bottom curve) for E$>$ 0.2 keV, in comparison.
Spectra were fitted with a {\it vmekal+sresc+gauss} model.}
\end{figure*}
The spectra (Fig. 7) show that both cameras EPIC-PN and EPIC-MOS1,2 detect the line at the same energy 
and at the same flux level within the flux uncertainties. The right-hand spectrum of Fig. 7 shows that
the line energy is the same for different regions (NW and S rims) of the RX J0852.0-4622 shell. The
continuum is best fitted by the {\it{sresc}} model, but is consistent also with a power law fit (see for more details Iyudin et al. 2004).
To illustrate the line excess Figure 8 shows a section of the spectrum between 3 and 6 keV; the left-hand spectrum shows 
the fit without a line whereas the fit shown in the right-hand side spectrum includes an emission line. 
\begin{figure*}[hbt]
  \centering
\includegraphics[bb=20 495 575 682,width=17.5cm,clip]{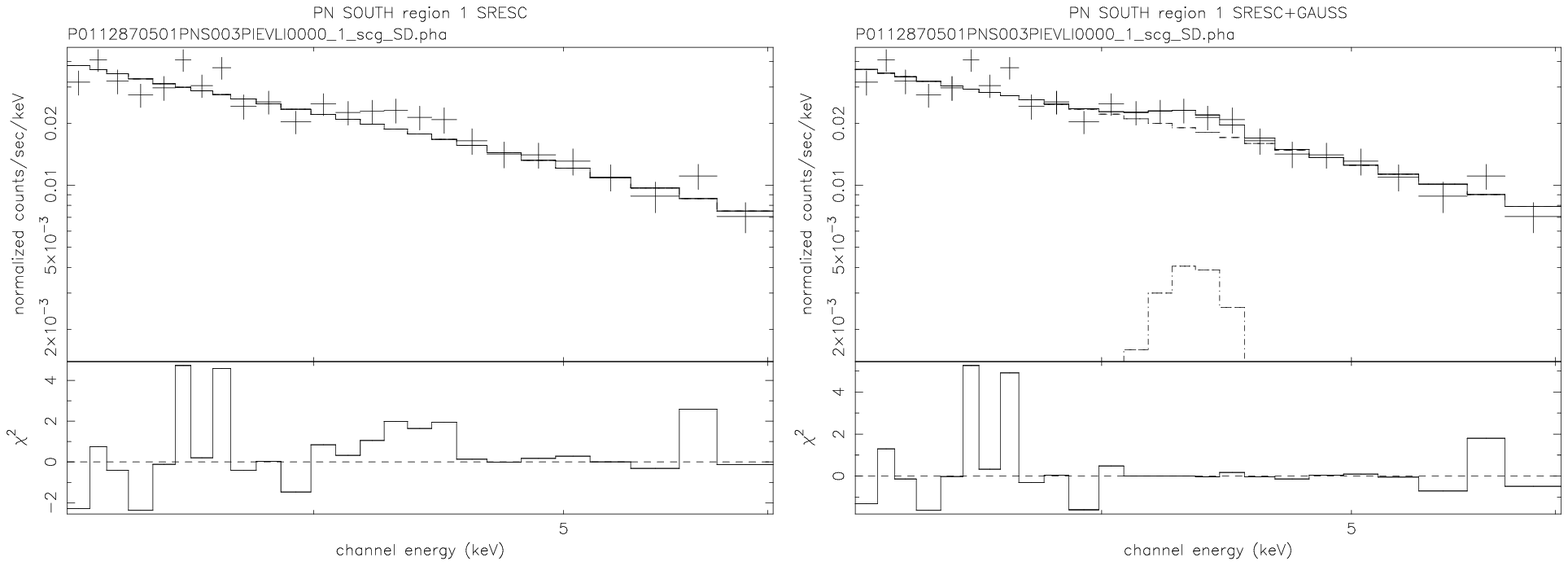}
  \caption{EPIC-PN spectrum of the RX J0852.0-4622 southern rim brightest region 1 for
3.0$<$E$<$ 6.0 keV; left: {\it sresc} model fit, right: the same spectrum fitted with a {\it sresc+gauss} model.}
\end{figure*}

\subsection{The western rim}

While the NW rim spectra are quite distinct from the underlying thermal emission of the Vela SNR that is dominating
at E$_x$$\leq$0.8 keV, the western rim spectra for E$_x$$\geq$0.8 keV show apart from a steep power law component
with a photon index of $\sim$4.7 
an additional 
thermal component, which is characterized by N$_H$$\sim$(2.84$\pm$0.15)$\times$10$^{21}$ cm$\sp{-2}$ and kT$\sim$0.2 keV.
Both the power law and the thermal component are significantly different from those of the NW and S rim. 
We note that the PN observation of the W rim was heavily contaminated by  solar proton flare events that 
lowered the 
effective exposure by a factor of three when the {\it{good time interval}} selection process 
was applied. The MOS spectra were
not strongly affected by the soft protons and confirm the line emission at 4.4$\pm$0.18 keV, also for the western rim.
\begin{figure*}[hbt]
  \centering
\includegraphics[bb=20 495 575 682,width=17.5cm,clip]{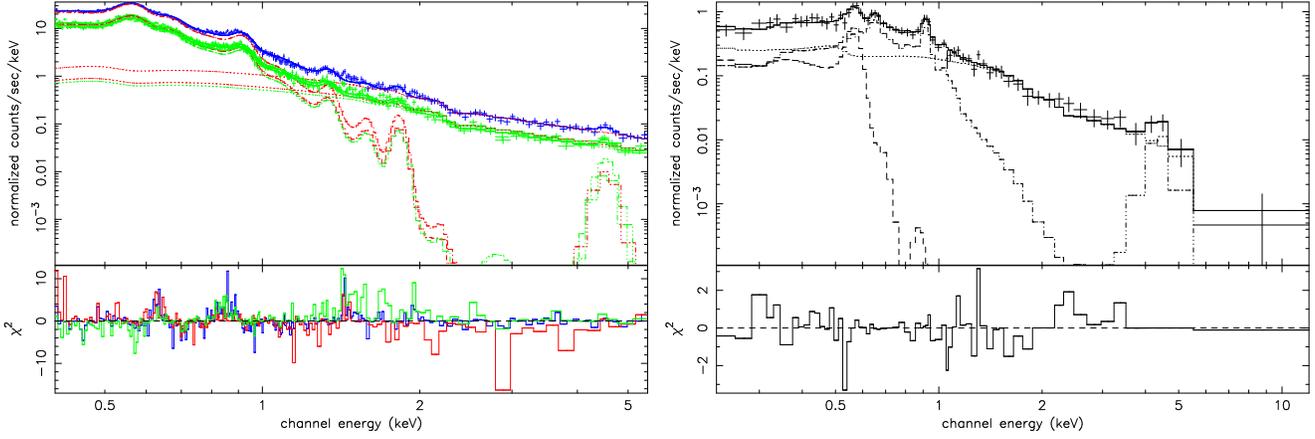}
  \caption{ Left: EPIC-PN spectra of the RX J0852.0-4622 western rim brightest regions,
for E$>$ 0.4 keV. Right: EPIC-MOS1 spectrum of the western rim region, that overlaps with the brightest regions of the EPIC-PN source region. Spectra were fitted with the {\it 2vmekal+powerlaw+gauss} model.}
\end{figure*}
\par
The western limb shows up  in the X-ray image of ROSAT as the X-ray bright feature W (Fig. 1).
We note that the radio emission from this part of RX J0852.0-4622 at 2420 MHz is rather weak.
This low  radio-continuum emission is consistent with the fit to the X-ray spectrum  (Fig. 9), which reveals 
a relatively small 
contribution of the synchrotron radiation in this region of the SNR.

\section{Discussion}

\subsection{Production mechanisms for the 4.2 keV emission line}

The emission line feature at $\sim$4.2 keV found by {\it{ASCA}} in the NW rim spectrum of RX J0852.0-4622 is also found 
in the {\it{XMM-Newton}} data (Fig. 4, 5, 7, 8 and 9). We have detected two line-like excesses for E$>$ 4 keV
in the region 1 of the NW rim (Fig. 4), but only one of these lines is visible in the spectra of the western and
southern rims.
This strongest line is at E$_{line}$=4.24$^{+0.18}_{-0.14}$ keV for the NW reg. 1 for a {\it power law}
continuum model, and E$_{line}$=4.44$\pm$0.11 keV for the same region using the {\it sresc} model. The
spectra of the western and southern rim show the line as well, at an energy which is consistent
with that of the NW rim. Our values of the line energy are slightly higher than the value given by Tsunemi
et al. (2000), but are consistent with the {\it{ASCA}} SIS spectrum of the NW rim (Fig. 4 in the paper by
Tsunemi et al. 2000). The line energies measured with {\it{XMM-Newton}} are summarized 
in Table 4. 

\begin{table*}[bht]
\begin{center}
\caption{Line feature parameters derived from the XMM spectra of the SNR rims.}
\begin{tabular*}{126mm}{||c|c|c|c|c|c|c||@{\extracolsep\fill}} \hline
&  & & & E$_{line}$&  F$_{line}$, 10$^{-6}$ & $N_H$ \\
\cline{5-7}
  \raisebox{1.5ex}[0pt]{Region} &  \raisebox{1.5ex}[0pt]{Obs ID} &  \raisebox{1.5ex}[0pt]{EPIC} &
 \raisebox{1.5ex}[0pt]{Model} & keV & ph cm$^{-2}$ s$^{-1}$ & (10$^{21}$ cm$^{-2}$) \\
\hline
\hline
NW rg1 &0112870301 &PN  & {\it plaw+vmek}  & 4.24$^{+0.18}_{-0.14}$ &  3.8$^{+3.5}_{-2.8}$   & 5.0\\
\hline
NW rg1 &0112870301 & PN  & {\it sresc}  & 4.44$\pm$0.11 &  4.24$^{+3.3}_{-3.2}$ & 3.9 \\
\hline
NW rg2 &0112870301 &PN  & {\it sresc}  & 4.2$\pm$0.2 &  2.4$^{+2.7}_{-1.5}$ & 3.2 \\
\hline
South &0112870501 & PN  & {\it sresc}  & 4.34$\pm$0.09 &  2.14$^{+2.6}_{-1.4}$ & 4.1 \\
\hline
West &0112870401&PN & {\it sresc}  & 4.54$\pm$0.07 &  9.5$^{+8.0}_{-5.0}$ & 2.8 \\
\hline
West &0112870401&MOS1 & {\it sresc}  & 4.43$\pm$0.18 &  26.0$^{+14.0}_{-14.0}$ & 2.6 \\
\hline
\hline
\end{tabular*}
\end{center}
\end{table*}

The complete data set of lines is consistent with one value 
for the line energy common to all three 
sections of the rim, which is  
$E_x = 4.45 \pm 0.05$ \hspace{2mm}keV.
\par

This line energy can be used for the ion transition identification. From the literature we have found
that the closest line is the Ly-$\beta$ line of H-like Sc, and the next nearest line is the 
Ly-$\alpha$ line of Ti (see Table 5, Deslattes et al. 2003 and references therein). 

\begin{table*}[bht]
\begin{center}
\caption{Positions, relative intensities of lines and K-shell fluorescent yields of
the abundant elements expected for the SNR RX J0852.0-4622 outer shell.}

\begin{tabular*}{130mm}{||l||l||c|c|c|c|c|c||@{\extracolsep\fill}} \hline
 & Transition & E$_x$$^1$, & E$_x$$^2$, & E$_x$$^3$, & E$_x$$^4$, & Relative & ${\omega}_k$$^5$,  \\

 \raisebox{1.5ex}[0pt]{Element} &  & (keV) &(keV) & (keV) & (keV)& imtensities $^1$ & recommended$^5$ \\
\hline
\hline
 $^{22}$Ti  & Ly${\alpha}_2$ & 4.5035  & 4.5049 & 4.50486 & 4.504 &0.339$\pm$0.040 & 0.219$\pm$0.018 \\
\hline
   & Ly${\alpha}_1$ & 4.5097  & 4.5109 & 4.51084 & 4.510 & 0.661$\pm$0.020 & \\
\hline
   & Ly${\beta}_{1,3}$ & 4.927  & 4.932 & 4.93181 & 4.931 & 0.248$\pm$0.032 & \\
\hline
 $^{21}$Sc  & Ly${\alpha}_2$ & ---  & --- & 4.0861& 4.085 & $\sim$0.34 & 0.190$\pm$0.016  \\
\hline
   & Ly${\alpha}_1$ & ---  & --- & 4.0906& 4.090  & $\sim$0.66 & \\
\hline
   & Ly${\beta}_{1,3}$ & ---  & --- & 4.4605 & 4.460& $\sim$0.15 & \\
\hline
 $^{20}$Ca  & Ly${\alpha}_2$ & 3.6890  & 3.6881 & 3.68809 &3.687 & 0.34$\pm$0.04 & 0.163$\pm$0.016 \\
\hline
   & Ly${\alpha}_1$ & 3.6900  & 3.6917 & 3.69168 & 3.691 &0.66$\pm$0.02 &  \\
\hline
   & Ly${\beta}_{1,3}$ & 4.0123  & 4.0125 & 4.0127 & 4.012 & 0.103$\pm$0.020 & \\
\hline
 \hline
\end{tabular*}
\end{center}
\par
$^1$ - Verma (2000), $^2$ - Sevier (1979), $^3$ - Bearden (1967), $^4$ - Johnson \& White (1970), $^5$ - Bambynek et al. (1972), 
the upper left indices attached to the element names in the most left-hand column corresponds to the nuclear charge.  
\end{table*}
It is still possible that the line is 
the He-like Sc line shifted to higher energy 
if the line has been produced in ion-atom collisions and has a few satellites (Watson et al. 1974). Depending on the line identification one can consider different production mechanisms.
\par
Assuming that the detected line is a H- or He-like line of $^{44}$Sc, that is produced via the $^{44}$Ti
decay by electron capture, we expect to detect a flux of $\sim$7.2$\times$10$^{-6}$ ph cm$^{-2}$ s$^{-1}$
from the whole SNR, based on the flux of (3.8$\pm$0.7)$\times$10$^{-5}$ ph cm$^{-2}$ s$^{-1}$ of the line at
1.157 MeV from the $^{44}$Ti decay (Iyudin et al. 1998). In fact we have detected a total line flux summed
over the three X-ray brightest rim sections of $\sim$1.83$\times$10$^{-5}$ ph cm$^{-2}$ s$^{-1}$, 
which exceeds the expected
flux but by less than 2 $\sigma$ 
with respect to the measured EPIC-PN  flux values and their uncertainties (see Table 4). 
\par
The electron loss or capture processes by ions moving in dilute gases are well studied (Betz 1972).
The equilibrium charge distribution of the SNR shell ions can be calculated based on the available
information about the element abundances in the SNR shell and those of the ISM. One can use the 
semi-empirical relations of Betz (1972) to compute the mean charge of ions moving in gases. 
Taking just two values for the expansion
velocity of the SNR shell, i.e. $\sim$5,000 km/s or 15,300 km/s (Iyudin et al. 1998) one
can derive from the empirical relation of Fig. 5.11 of Betz (1972) the mean charge of the 
Sc ions which move with the SNR shell to $<q>$=$\sim$4.4 and $\sim$13.2, respectively.  
 To describe the charge distribution of the ions moving in 
dilute gaseous matter we use the approach of Nikolaev (1965) and of Betz (1972) again. In
the first approximation the experimentally derived equilibrium charge distribution of ions in
gaseous matter is well described by a Gaussian with the r.m.s. value $d$ that can be expressed
as $d=0.27 Z^{1/2}$, and by the mean charge of all ions of a particular element of nuclear charge 
 $Z$. The equilibrium charge distribution can be written as
\begin{equation}
F_q \approx \left(2 \pi d^2\right)^{-1/2} exp [-\left(q-<q>\right)^2/2 d^2].
\end{equation}
From this Gaussian distribution it is obvious that even for the highest derived expansion velocity
of the SNR shell (Iyudin et al. 1998) the equilibrium charge distributions of $^{44}$Ti and the
$^{44}$Sc will be limited to the range of charges $\leq$16 with a probability of $\sim$99\%.
For solar abundances of the ISM the influence of the known asymmetry in the equilibrium charge distribution 
 created by high Z gas mixtures has a minor effect as far as the extension of the charge
distribution towards higher ionization states is concerned. For the target gas consisting mainly of
hydrogen and helium the equilibrium charge distribution remains Gaussian (symmetric) (Betz 1972).
It means that without invoking other inner-shell ionization mechanisms we should not have
hydrogen- or helium-like transitions in Sc or Ti in  the SNR shell.
\par
Note that the situation is completely different for the Cas A SNR where thermal processes dominate the 
formation of the charge distribution, in  contrast to the case of RX J0852.0-4622 where the
reverse shock is likely to be still very weak (Iyudin \& Aschenbach 2001).

\subsection{Charge states of the abundant ions in the outer rim}

To estimate the line flux two hypotheses for the origin seem plausible at this point in time, i.e., (1) - the line
is the result of  inner-shell ionization of the abundant
constituents of the SNR rim 
 induced by electron capture decay
  , or (2) - the line is a consequence of collisions of ions present in the SNR outer 
 shell with the ambient matter, predominantly with the ISM hydrogen atoms or electrons.
In the latter case the width and the shift of the line are carrying information about the
expansion velocity of the SNR (Moore et al. 1972, Kauffman et al. 1973, Garcia et al. 1973, Verma 2000,
Iyudin and Aschenbach 2001).
Most likely the line at $\sim$4.4 keV is  a blend of lines
emitted by $^{44}$Sc after $^{44}$Ti decay. But one cannot a priori exclude the possibility that at least
part of the line flux is produced via the inner-shell ionization of Ti induced by  collisions with
 ambient matter. 
\par
Two Ti isotopes are abundantly produced  and transported to the outer layers of the
sub-Chandrasekhar type Ia SNe ejecta, which  are $^{44}$Ti and $^{48}$Ti, the latter is originally
produced as $^{48}$Cr in quantities comparable to $^{44}$Ti (Livne \&  Arnett 1995, Woosley \& 
Weaver, 1994). Interestingly, $^{48}$Cr decays via electron capture (100\%) to $^{48}$V, which
in turn decays to $^{48}$Ti via electron capture (50.4\%) and ${\beta}^+$ decay (49.6\%). This sequence
will eventually leave the $^{48}$Ti isotope also in ionized and excited states. The life-times
of $^{48}$Cr and $^{48}$V are 21.56 hr and 15.976 days, respectively. Therefore,
the K-shell X-ray transitions following the Cr and V decays via electron capture have happened long ago,
and no bright X-ray lines are expected from these two isotopes unless further inner-shell vacations are 
generated by ion interactions with the ISM.
The consequence of the K-shell vacancies after the inner-shell ionization in the collisions with the
ambient atoms is that predominantly K-$\alpha$ and/or K-$\beta$ lines of Ca and Ti are emitted, with
relative line strengths of K-${\alpha}$ and K-${\beta}$ given by the corresponding fluorescence yields 
 (c.f. Table 5).
\par
The scenarios introduced above can also serve to answer questions about the exclusive 
inner-shell ionization of just Sc and Ti. 
Given the X-ray brightness distribution of RX J0852.0-4622 it seems that this SNR only has started 
to interact with denser ISM matter at the north-western and southern boundaries, if at all. 
There is no clear indication of the existence of a reverse shock, which would be the consequence 
of such an interaction. Only the presence of the two filament-like structures in the north-western 
section may be taken as evidence for a reverse shock and if so it only recently has started to develop. 
If RX J0852.0-4622 is indeed young
(Iyudin et al. 1998, Aschenbach et al. 1999), then its expansion velocity can still be as high as
indicated by the width of the $\gamma$-ray line (Iyudin et al. 1998), e.g. V$_{exp}$$\sim$15.000 km s$^{-1}$.
In this case ions contained in the outermost layers of the ejecta will have kinetic energies of $\sim$1
MeV/amu, and therefore will experience internal ionisation of the K-shell through collisions with the heavy
atoms (ions) of the ambient matter, possibly  contained in clouds. As a result of such an interaction also the 
emitted K-$\alpha$ and K-$\beta$ lines of Sc and Ti
could be broadened and shifted to higher energies because of the contribution of  satellite lines created in these
heavy ion-atom collisions (Moore et al. 1972, Kaufmann et al. 1973, Whalen \& Briancon 1975). But such 
an interpretation seems to be in conflict with measured  
position, shape and energy spread of the X-ray lines.
\par
The lines detected by the {\it{XMM-Newton}} EPIC-PN camera in each of the  observed three sections of the SNR 
seem to be slightly broadened with an r.m.s. width of  150 eV, that has to be compared with  
the instrumental energy resolution of $\leq$125
eV at the appropriate energy of ~4.4 keV (EPIC-PN, Haberl et al. 2002). 
The natural widths of the  K-$\alpha$ lines of elements with Z$\sim$20 
are $\leq$2 eV (Krause \& Oliver 1979). The energy shift of the line position depends 
 on the projectile type and its kinetic energy (Moore et al. 1972, Kauffman et al. 1973, Verma 2000),
and has a maximum value of $\sim$80 eV for Ti (Moore et al. 1972, Verma 2000). But an energy shift of the detected line cannot be unambigously claimed from the data. Instead, we believe that the line position 
is consistent with the unshifted K-$\beta$ line of Sc.
At last the line broadening of $\sim$80 eV could be  produced by Doppler motion, that would lead to
a radial projected average velocity of the emitting volume of the order of $\sim$5800 km/s. This
value for the expansion velocity of RX J0852.0-4622 is lower than the value derived from the broadening of
the line at 1.157 MeV (Iyudin et al. 1998). But this X-ray derived value may not be in a conflict with the $\gamma$-ray derived velocity, given the possibility that the 
geometry and the velocity of the emitting volume elements need not to be the same.

\subsection{GRO J0852-4642 spectrum and progenitor}

In the following section we take the view that the measurements of $^{44}$Ti with {\it{COMPTEL}} are real and 
not a statistical excursion in the spectrum (Iyudin et al. 1998). Both the {\it{ASCA}} and the {\it{XMM-Newton}} 
measurements of the 4.4 keV emission feature support this interpretation, and we use this information to 
access the type of progenitor of RX J0852.0-4622 / GRO J0852-4642. 
 $^{44}$Ti is believed to be synthesized near the mass cut interface in core-collapse SNe explosions 
(Woosley and Weaver 1995, Timmes et al. 1996).
Therefore, the total yield and spatial distribution of $^{44}$Ti produced in such an explosion are very
sensitive to the explosion mechanism and the ejecta dynamics, e.g. in SNe of type Ia   a mass cut
does not exist. 
 Assuming that the $^{44}$Ti line profile measurements by {\it{COMPTEL}} (Iyudin et al. 1998) are not far
from reality, which is supported by comparison with the $^{26}$Al line shape derived for the same instrument (see Fig. 2
in Iyudin et al. 1998), we need to understand expansion velocities of 15300$\pm$3700 km s$^{-1}$
for $^{44}$Ti, and $\leq$8000 km s$^{-1}$ for $^{26}$Al.
We compare the measurements of both the expansion velocity and the abundances of Ca and Ti with the 
model predictions of Woosley and Weaver (1994) and Livne and Arnett
(1995). It turns out that our measurements are only met by models 
 for a sub-Chandrasekhar type Ia supernova explosion. This means that the progenitor of 
 RX J0852.0-4622 was very likely a 0.6-0.7 M$_{\odot}$ helium accreting white dwarf and according to present knowledge 
no compact remnant is expected to be associated with RX J0852.0-4622.
Of course we are then left with the question about the nature and origin of the central point-like source 
CXOU J085201.4-461753.  
Clearly, the ionization state of the Ti can have an impact on the above conclusion but of moderate size.
For example, values of age and distance to the SNR may be underestimated, but not more than by $\sim$30\% 
(Aschenbach et al. 1999). Mochizuki et al. (1999) have modelled the heating and ionization of $\sp{44}$Ti
by the reverse shock in Cas A, for which they report the possibility of  a currently
 increased $\sp{44}$Ti activity.
With respect to RX J0852.0-4622 / GRO J0852-4642 they find that the reverse shock does not heat
the ejecta to sufficiently high temperatures to ionize $\sp{44}$Ti because of the low ambient
matter density.

\section{Conclusion}

{\it{XMM-Newton}} observations of three sections on the rim of RX J0852.0-4622 have been carried out. The north-western rim has been 
resolved in two clearly separated  filament-like structures. The high energy section of the spectra of
 both the north-western  rim and the southern rim 
are consistent with a power law shape and a power law which shows a roll-off towards high energies. Interestingly, the roll-off energy of about 
1 keV is fairly high and it remains an open question what the spectrum looks like towards lower energies. The difficulty here is 
the contribution from the Vela SNR, which is fairly bright at energies below 1 keV. The various models seem to verify an absorbing 
column density towards RX J0852.0-4622 which is significantly higher than typical for the Vela SNR. Whether this implies a greater distance 
is not clear. The spectral slope measured with {\it{XMM-Newton}} is basically the same as what has been measured with {\it{ASCA}}. 
The western part of the remnant is different as it shows only a small contribution by a power law and if present at all the 
spectrum is much steeper. The {\it{ASCA}} measurements indicated the presence of an emission line like feature at around 4.1 - 4.2 keV. 
The {\it{XMM-Newton}} data confirm such a feature, and it seems to be present everywhere on the remnant's rim. The line energy 
averaged over the three observational fields is 4.45 $\pm$ 0.05 keV. We attribute this line or lines to the emission of Ti and Sc which might be 
excited by atom/ion or ion/ion collisions. 
The X-ray line flux expected from such an interaction is consistent with the 1.157 MeV $\gamma$-ray line flux measured 
by {\it{COMPTEL}}. This consistency of the X-ray line flux and the $\gamma$-ray line flux 
lends further support to the existence and amounts of Ti in RX J0852.0-4622 claimed by Iyudin et al. (1998) and to the suggestion 
that RX J0852.0-4622 
is young and nearby (Aschenbach et al. 1999). 
Iyudin et al. (1998) quote a very large broadening of the 1.157 MeV $\gamma$-ray line which would indicate a large 
velocity of the emitting matter of about 15.000 km/s. Such a high ejecta velocity for Ti is found only in explosion models of 
sub-Chandrasekhar type Ia supernovae (Woosley \& Weaver 1994, Livne \& Arnett 1995).        
In this case no compact remnant is expected. The obvious questions remaining are what the nature and the origin of the 
central compact source CXOU J085201.4-461753 are and why the absorption column density apparently associated with RX J0852.0-4622 is much 
greater than typical for Vela. 

\section{Acknowledgements}

The authors acknowledge the comments of the anonymous referee that helped to improve the quality of the paper. The XMM-Newton project is an ESA Science Mission with instruments and
contributions directly funded by ESA Member States and the USA (NASA). The
XMM-Newton project is supported by the Bundesministerium f\"ur Bildung und
For\-schung / Deutsches Zentrum f\"ur Luft- und Raumfahrt (BMBF / DLR), the
Max-Planck-Gesellschaft and the Heidenhain-Stiftung. The COMPTEL project was supported 
by the BMBF through DLR grant 50 QV 9096 8.
AFI acknowledges financial support from the BMBF through the DLR grant 50 OR 0002.

\end{document}